\documentclass[preprint]{aastex}
\usepackage{graphicx}
\newcommand{\hi}{\protect\ion{H}{1}}

\shorttitle{Radio AGN in Fossil Groups}
\shortauthors{Hess et al.}

\begin{document}

\title{Fresh Activity in Old Systems:\\
  Radio AGN in Fossil Groups of Galaxies}
\author{Kelley M. Hess\altaffilmark{1,2}, Eric M. Wilcots\altaffilmark{2}, \& Victoria L. Hartwick\altaffilmark{2}}

\altaffiltext{1}{University of Cape Town,
  Department of Astronomy, Private Bag X3, Rondebosch 7701\\ \emph{e-mail:} hess@ast.uct.ac.za}
\altaffiltext{2}{Department of Astronomy,
  University of Wisconsin-Madison, Madison, WI 53706 \\ \emph{e-mail:} ewilcots@astro.wisc.edu, vhartwick@wisc.edu}

\begin{abstract}
We present the first systematic 1.4 GHz Very Large Array radio continuum survey of fossil galaxy group candidates.  These are virialized systems believed to have assembled over a gigayear in the past through the merging of galaxy group members into a single, isolated, massive elliptical galaxy and featuring an extended hot X-ray halo.  We use new photometric and spectroscopic data from SDSS Data Release 7 to determine that three of the candidates are clearly not fossil groups.  Of the remaining 30 candidates, 67\% contain a radio-loud (L$_{1.4GHz} > 10^{23}$ W Hz$^{-1}$) active galactic nucleus (AGN) at the center of their dominant elliptical galaxy.
We find a weak correlation between the radio luminosity of the AGN and the X-ray luminosity of the halo suggesting that the AGN contributes to energy deposition into the intragroup medium.  We only find a correlation between the radio and optical luminosity of the central elliptical galaxy when we include X-ray selected, elliptically dominated non-fossil groups, indicating a weak relationship between AGN strength and the mass assembly history of the groups.  The dominant elliptical galaxy of fossil groups is on average roughly an order of magnitude more luminous than normal group elliptical galaxies in optical, X-ray, and radio luminosities and our findings are consistent with previous results that the radio-loud fraction in elliptical galaxies is linked to the stellar mass of a population.
The current level of activity in fossil groups suggests that AGN fueling continues long after the last major merger.  We discuss several possibilities for fueling the AGN at the present epoch.
\end{abstract}


\section{Introduction}

A ``fossil galaxy group'' is believed to be the merger remnant of a galaxy group whose most massive members have succumbed to dynamical friction.  The low velocity dispersion of the group environment results in a high interaction rate between galaxies, such that the most massive members eventually coalesce to form a single giant elliptical.  What remains is the faint dwarf galaxy population, and an extended, hot gaseous halo that fills the dark matter potential of the former group.  As the dynamical end point of groups, fossil groups appear to be undisturbed, virialized systems.  They are identified observationally by two criteria: fossil groups are dominated by a single elliptical galaxy, typically brighter than L$_*$, where the next brightest companion is fainter by at least two magnitudes in R-band; and an extended, hot X-ray halo, typically brighter than L$_X=10^{42}$ ergs s$^{-1}$ \citep{Jones03}. 

The first fossil group system, RX J13040.6+4018, was identified by comparing ROSAT extended deep survey images with Palomar sky survey plates \citep{Ponman94}.  In the last decade, several authors have contributed to a growing list of known fossil groups (e.g. 3, \citealt{Vikhlinin99}; 5, \citealt{Jones03}; 3, \citealt{2004AdSpR..34.2525Y}; 1, \citealt{2004MNRAS.349.1240K}; 1, \citealt{Sun04}; 1, \citealt{2005ApJ...624..124U}).  Recently, \cite{Santos07} used the Sloan Digital Sky Survey (SDSS) Data Release 5 (DR5) and the ROSAT All-Sky Survey (RASS) to catalog 34 fossil group candidates allowing for the beginning of statistical investigations into the nature of such objects. Since the undertaking of our radio survey, \cite{Eigenthaler09} reported 34 new fossil groups, and \cite{LaBarbera09} identified an additional 25 fossil group candidates in SDSS DR4 and RASS.

The formation of fossil groups is inferred from the comparison of the optical luminosity of the central elliptical galaxy and the X-ray luminosity of the halo with observations of systems in a range of similar or intermediate evolutionary states.  The optical and X-ray values of fossil groups are comparable to compact galaxy groups---systems believed to be in the most advanced stages of merging \citep{MendesdeOliveira07}.  Further, they are more numerous than compact groups suggesting that the merging occurs on short timescales compared to the lifetime of the group (e.g. \citealt{Mamon87}).  The end result is the largest galaxies in compact groups succumb to dynamical friction to coalesce into a single, massive elliptical galaxy \citep{MendesdeOliveira06,MendesdeOliveira07}.  

Cosmological N-body simulations suggest that fossil groups formed early in the lifetime of the Universe, assembling half their mass by redshift of $z\ge1$ \citep{D'onghia05}, or by $z\ge0.8$ \citep{vonBendaBeckmann08}.  Observationally, the luminosity function, the large magnitude gap between the first and second ranked galaxies, and the large virialized X-ray halo are believed to be indicative of this early formation \citep{Ponman94,Jones03,MendesdeOliveira06}.  However, simulations also show that fossil groups may only be a phase of hierarchical structure evolution which ends with the fresh infall of massive galaxies from the surrounding environment \citep{vonBendaBeckmann08,Dariush10}.  It has been noted that the dominant elliptical galaxy in fossil groups frequently resembles that of the brightest cluster galaxies (e.g. \citealt{Vikhlinin99,MendesdeOliveira06}), and because fossil groups have a similar space density to galaxy clusters ($\sim$$10^{-6}$ Mpc$^{-3}$; \citealt{Jones03}; \citealt{Santos07}), it has been suggested that may be the early seeds around which present-day clusters form \citep{Jones03}.

Fossil groups are most well studied at optical and X-ray wavelengths, and it has been demonstrated that a strong relationship exists between the X-ray and R-band luminosities of the hot halo and the central dominant elliptical galaxy, respectively. This relationship indicates that the conditions of the group environment are linked to the mass assembly history of the fossil group \citep{Jones03}.  However, it is unclear what mechanism is responsible for elevating the X-ray temperature in fossil groups.

Clusters and elliptically dominated non-fossil groups follow the well known X-ray luminosity-temperature relationship (e.g.~\citealt{White97,Jones00}) and the temperatures of these systems imply the presence of non-gravitational heating in the intracluster and intragroup medium (ICM/IGM).  Radio active galactic nuclei (AGN) commonly occur in these systems and their detailed morphology frequently corresponds to cavities in the X-ray halo, thus they are invoked to account for  excess energy in the ICM/IGM (e.g. \citealt{Birzan04,Wise07}).  
Fossil group X-ray halos for which the temperature has been measured appear to follow a similar $L_X-T_X$ relation  although they are more luminous relative to non-fossils implying that if the X-ray luminosity is boosted, so is the temperature \citep{Khosroshahi07}. Among non-fossil groups, evidence shows that radio-loud groups are hotter in X-rays than radio-quiet groups, indicating that on-going nuclear activity contributes to heating the IGM \citep{Croston05}.  
Simulations show that radio jets from AGN may deposit significant amounts of energy in the IGM (e.g. \citealt{Nath02,Heinz06}), and both simulations and observations of clusters support the theory that radio AGN can provide a sufficiently distributed heating mechanism to account for the excess energy (\citealt{Bruggen02,Fabian03}).  
However, simulations by \citet{Rowley04} suggest that major mergers can also elevate the X-ray temperature and luminosity in clusters of a few $\times10^{14}$ M$_{\odot}$ and that this may last for up to a Gyr after the stellar merger signatures have faded, decreasing the necessity of AGN to account for excess energy in the IGM.  


A radio investigation is therefore important for understanding the role of radio AGN in heating the IGM in fossil groups.  A multi-wavelength investigation will allow us to link the properties of the radio AGN (or lack thereof) to the group environment as traced by the X-ray halo, and mass assembly of the central galaxy as indicated by the $r$-band stellar luminosity.
Further, if radio AGN are common in fossil groups, what physical mechanism is responsible for maintaining or triggering them long after the majority of mass assembly has taken place in these systems?
There is some evidence that optical AGN activity is the result of major mergers at late times \citep{Alonso07}, however observations and simulations suggest the last major merger in fossil groups occurred well over 1 Gyr in the past \citep{D'onghia05,vonBendaBeckmann08,Jones00,2006MNRAS.372L..68K}.  Additionally, radio AGN activity, or indeed any AGN activity, appears to be well removed from merger activity, particularly since a redshift of $z=1$ \citep{Schawinski10,Cisternas11}.
The thermodynamic properties of the X-ray halo gas, undisturbed by recent mergers, exhibited by the smooth, symmetric distribution of the X-ray contours, make the fossil group an ideal observational laboratory in which to test theories of AGN feedback for regulating the temperature of group and cluster sized halos \citep{Jetha09}.   

In this paper, we present the first systematic radio study of fossil group candidates taken from the \citet{Santos07} ROSAT/SDSS DR5 catalog, with a modestly deep 1.4 GHz Very Large Array (VLA) survey, and use our results to conduct a multi-wavelength investigation of fossil groups.  We measure the radio flux as a tracer of AGN activity and our results reveal the degree to which radio AGN are \textit{currently} contributing to the energy deposition of the IGM.  We calculate the fraction of radio-loud AGN and compare the population of radio sources with those in elliptically dominated groups and clusters which represent the continuum of hierarchical structures.  The presence of radio AGN indicate that physical mechanisms such as cooling flows, minor mergers, or late-time accretion, are important for sustaining an AGN late in the evolution of fossil group systems, long after the majority of mass assembly has taken place.  Thus, our radio survey is strongly motivated by outstanding questions in the evolution of galaxies, gas accretion, and the hierarchical formation of large scale structure.  

The rest of the paper is organized as follows: in Section \ref{obs} we describe our VLA 1.4 GHz radio observations and the data reduction, and describe the acquisition of optical data from SDSS DR7.  In Section \ref{results} we compare published X-ray data based on the \citet{Santos07} sample, with SDSS DR7 \emph{r}-band photometry and with our 1.4 GHz radio results.  In Section \ref{discussion} we discuss the possibilities for fueling a radio source long after the last major merger with the central, dominant elliptical galaxy.  Finally, we conclude in Section\ref{conclusions}.

\section{Sample Selection, Observations, and Data Reduction} \label{obs}

Our radio survey is derived from the catalog of 34 fossil group candidates identified by \citet{Santos07}.  The catalog was constructed by comparing \emph{r}-band photometry from Sloan Digital Sky Survey (SDSS) Data Release 5 (DR5) with X-ray observations from the ROSAT All-Sky Survey.  \citet{Santos07} followed the traditional definition of a fossil group, but loosened the X-ray criterion.  In this catalog a system qualified as a fossil group candidate when (1) the optical image consists of an elliptical galaxy where the difference in brightness between it and the next brightest galaxy is greater than 2 magnitudes in $r$-band, and (2) the candidate system is detected as an extended X-ray source.  The search volume was fixed to a radius of $0.5 h_{70}^{-1}$ Mpc from the dominant elliptical galaxy and a $\delta z$ equivalent to the SDSS photometric redshift uncertainty.  (For more discussion see Section 2 and 3.3 of \citealt{Santos07}).  Unlike previous searches for fossil groups, \citet{Santos07} do not impose a lower limit on the X-ray luminosity and do not assume a lower limit for the number of companion galaxies.  Nonetheless, they only find one system which lies below the typical X-ray cutoff of L$_X=10^{42}$ ergs s$^{-1}$ and they find no systems of isolated ellipticals.

\subsection{Radio Observations}

We observed 33 of the 34 fossil groups with the Very Large Array (VLA) in B- or BnC-configuration in Spring 2009 (Project Code AH0988).  Observations were conducted with a standard VLA L-band continuum mode of 8 channels and 25 MHz bandwidth.  We spent 50-60 minutes of total integration time centered on the dominant elliptical galaxy in each fossil group.  We observed a bright flux calibrator once at the beginning or end of each observing run and then alternated between our target and a phase calibrator every $\sim$20-25 minutes.  The fossil groups span an average of $2-5$ arcminutes on the sky, the largest spanning 13 arcminutes, which fits well within the 30 arcminute primary beam of the VLA at 1.4 GHz.  

The observations were reduced in the standard way using AIPS.  In imaging and deconvolution, we were careful to box and clean every source we could identify in the field in order to minimize flux contamination from sidelobes and we used faceting techniques to image bright sources far from the pointing center.  The resolution of the synthesized beam at 1.4 GHz in B-array ranges from 30 kpc at z = 0.489 to 2.5 kpc at z = 0.032, which allowed us to resolve structure in some of the sources.  In all fossil group detections, the radio source is associated with the central, dominant elliptical galaxy.  In one case, J104302.57+005418.2, a radio source detected near the pointing center is associated with a background galaxy.

We attain an \textit{rms} noise of $30-94~\mu$Jy beam$^{-1}$ for each detection.  For undetected sources we reach a mean \textit{rms} of $87~\mu$Jy beam$^{-1}$, 5.2 times deeper than NVSS \citep{1998AJ....115.1693C} and 1.7 times deeper than FIRST \citep{1995ApJ...450..559B}.  The \textit{rms} of our observations were in some cases limited by the presence of a nearby strong continuum source.

\subsection{Update of Optical Data}

To put our 1.4 GHz results in the context of previously published observations and simulations, we sought to calculate the difference in $r$-band magnitude between the first- and second-ranked galaxies of each fossil group ($\Delta$mag).  To find this quantity, we repeated the neighboring galaxy query described in Appendix A of \citet{Santos07} in the SDSS query database, CasJobs, on DR7.  The initial DR5 based catalog relied on photometric redshifts to identify fossil group candidates; in DR7 all first-ranked galaxies now have spectroscopic redshifts although the majority of neighbors are still photometric. 

The DR7 magnitudes reveal that six candidates no longer qualify as fossil groups, as they fail the optical criteria of having $\Delta$mag $< 2.0$.  Three of these candidates fail the criteria dramatically by having a $\Delta$mag $< 1.3$ between the first- and second-ranked galaxies.  The remaining three candidates are on the cusp of the fossil group criteria having $1.8 < \Delta$mag $< 2.0$.  However, as several studies have pointed out, $\Delta$mag $< 2.0$ is an arbitrary distinction by which to identify fossil groups.  In fact, fossil groups may be a short lived phase of evolution that is terminated with renewed infall from the surrounding environment \citep{vonBendaBeckmann08}.  Simulations demonstrate that alternate criteria may select older systems that are longer lived fossil groups \citep{Dariush10}, however there is clearly a continuum of structures.  As a result, we choose to include fossil group candidates with $\Delta$mag $> 1.8$ in our subsequent analysis, while dropping those with $\Delta$mag $< 1.3$.  Finally, we visually inspected the SDSS images and find that even with complete spectroscopic coverage it is unlikely any other candidates would further fail the optical fossil group criteria. The results from DR7 brings our sample to 30 remaining fossil groups for which we presently drop the ``candidate'' status.

\section{Results} \label{results}

We detect 20 (67\%) of the fossil groups in our VLA observations above the 3$\sigma$ level.  All but one of these sources are radio-loud (L$_{1.4GHz} > 10^{23}$ W Hz$^{-1}$) within their error bars.  The definition of ``radio-loud'' varies throughout the literature, however we chose this cutoff based on the findings of \citet{Sadler02} that above this luminosity the radio source population is dominated by AGN, while below it radio sources are primarily starbursting galaxies.

\begin{figure*}[t]
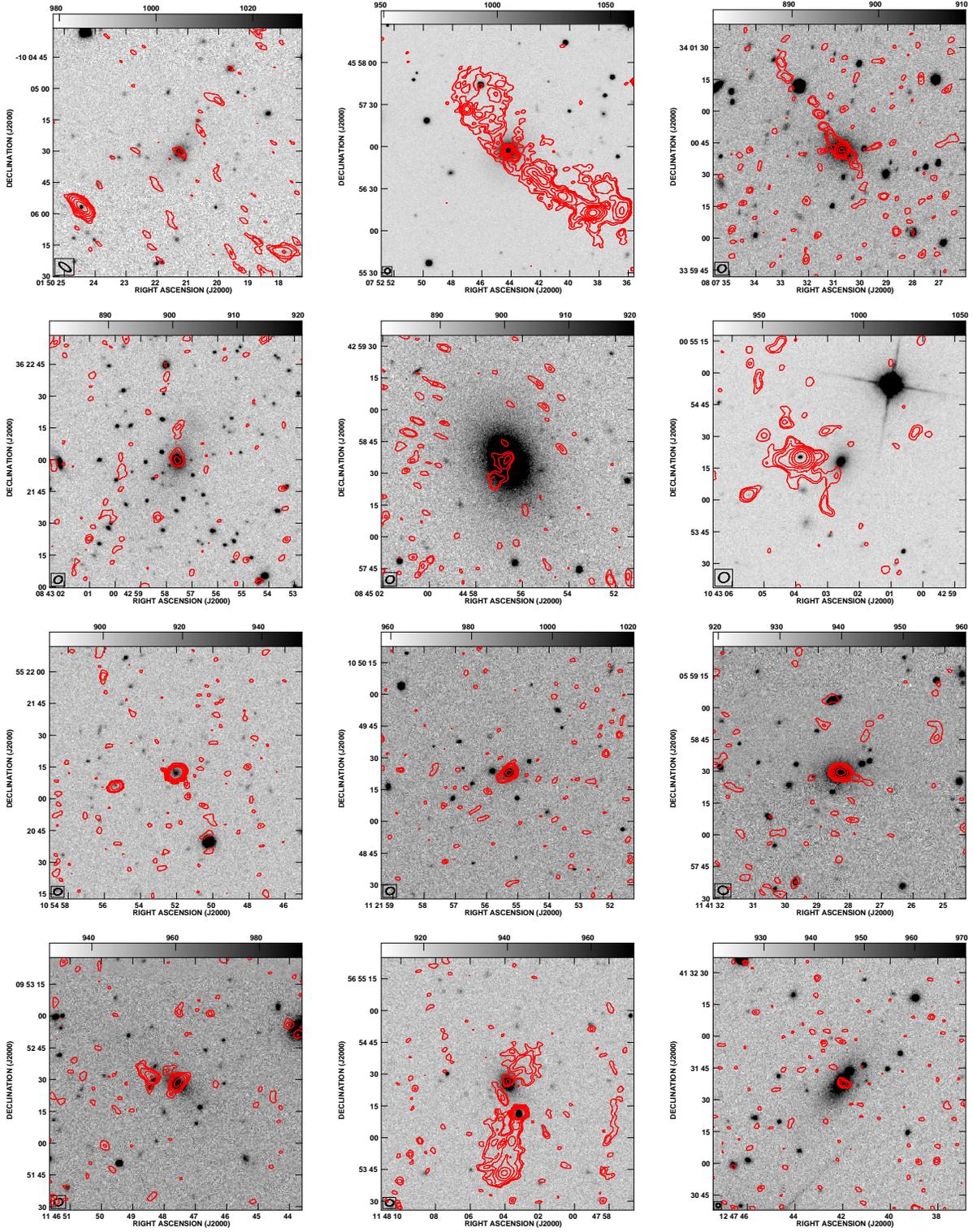

\begin{center}
$\begin{array}{ccc}
\includegraphics[width=2.0in]{f1a.eps} & \includegraphics[width=2.0in]{f1b.eps} & \includegraphics[width=2.0in]{f1c.eps} \\ 
\includegraphics[width=2.0in]{f1d.eps} & \includegraphics[width=2.0in]{f1e.eps} & \includegraphics[width=2.0in]{f1f.eps} \\ 
\includegraphics[width=2.0in]{f1g.eps} & \includegraphics[width=2.0in]{f1h.eps} & \includegraphics[width=2.0in]{f1i.eps} \\ 
\includegraphics[width=2.0in]{f1j.eps} & \includegraphics[width=2.0in]{f1k.eps} & \includegraphics[width=2.0in]{f1l.eps} \\
\end{array}$
\end{center}
\caption{A collage of fossil group candidates with radio detections at 1.4 GHz.  Radio contours are overlaid on SDSS \emph{r}-band images.  All but one of these sources are radio-loud ($L_{1.4GHz} > 10^{23}$ W Hz$^{-1}$) within the error bars. Contours are at 2, 3, 5, 10, 20, 40, 80$\sigma$ for each pointing, where $\sigma$ is the $rms$ noise given in Table \ref{data}. Panel numbers from Column 10 count left to right, top to bottom. Radio emission in panel 6 is associated with a background galaxy. The southern component of radio emission in panel 11 is associated with a background quasar at $z=0.451$}
\label{images1}
\end{figure*}

\begin{figure*}[t]
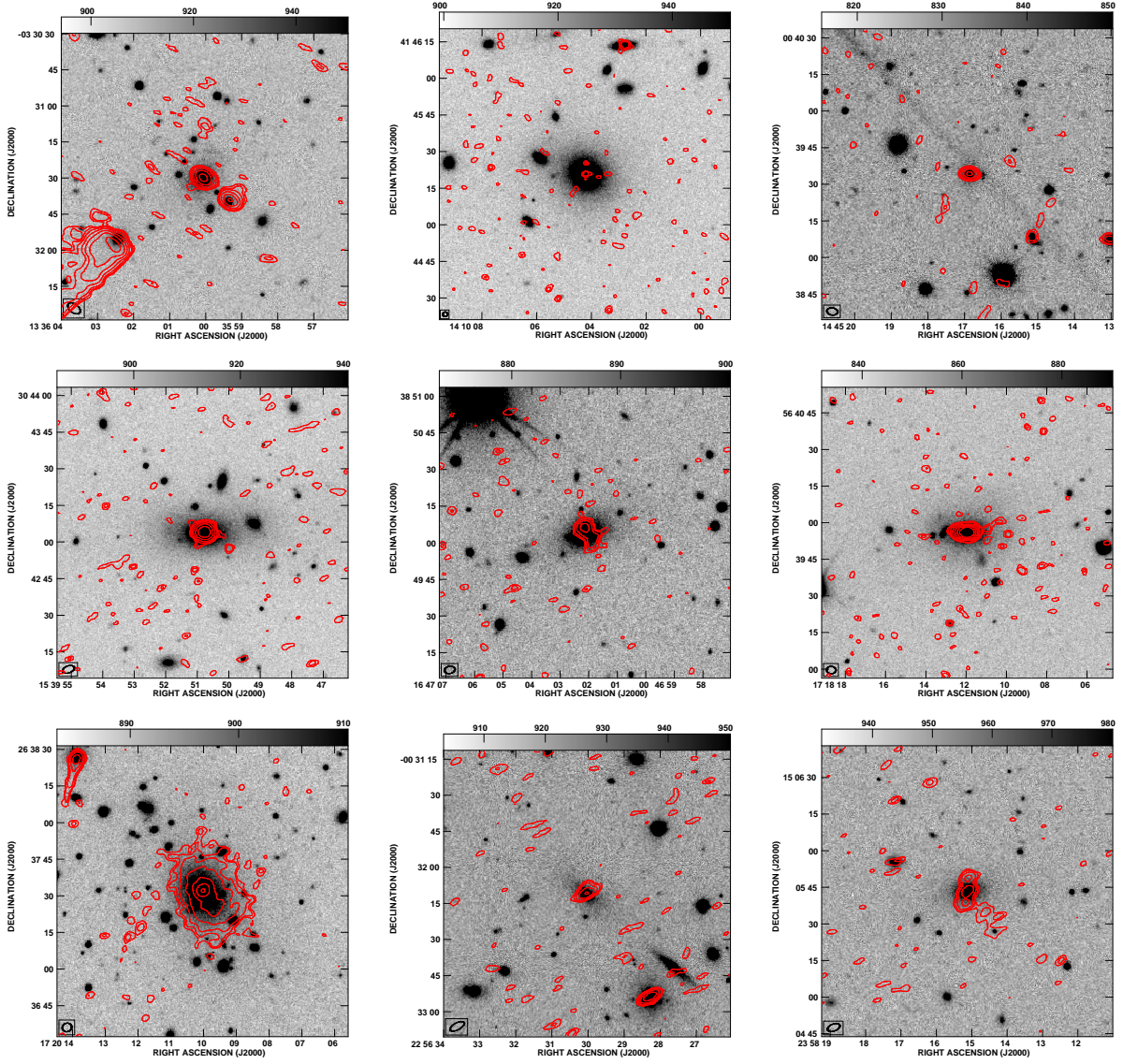

\begin{center}
$\begin{array}{ccc}
\includegraphics[width=2.0in]{f2a.eps} & \includegraphics[width=2.0in]{f2b.eps} & \includegraphics[width=2.0in]{f2c.eps} \\ 
\includegraphics[width=2.0in]{f2d.eps} & \includegraphics[width=2.0in]{f2e.eps} & \includegraphics[width=2.0in]{f2f.eps} \\
\includegraphics[width=2.0in]{f2g.eps} & \includegraphics[width=2.0in]{f2h.eps} & \includegraphics[width=2.0in]{f2i.eps} \\
\end{array}$
\end{center}
\caption{Same as Figure \ref{images1}.  Panel numbers referenced in Column 10 of Table \ref{data} go from 13 to 21.}
\label{images2}
\end{figure*}

In Figures \ref{images1} \& \ref{images2} we present the 1.4 GHz radio contours overlaid on SDSS \emph{r}-band images.  We believe these detections are unambiguous.  Given the color, morphology, and DR7 spectra of the central elliptical galaxies, we confirm that all of our radio sources are associated with AGN activity.  Roughly half of the radio detections are unresolved point sources while half show extended emission and sometimes detailed structure ranging from fuzzy, extended halos to dramatic bipolar jets with large lobes.  In the sole case of an off-center extended radio source, J104302.57+005418.2, it is clearly associated with a background galaxy (panel 6 of Figure \ref{images1}).  In the case of J114803.81+565425.6 the brightest radio source is located south of the pointing center and is associated with a background quasar at a redshift of $z=0.451$ (panel 11 of Figure \ref{images1}).  Therefore, we only consider emission located at the position of the central elliptical galaxy and extending continuously to the northwest as emission associated with the fossil group. Only five of the \citet{Santos07} candidates were previously detected in NVSS or FIRST demonstrating the necessity of deeper radio observations for multi-wavelength studies of fossil group candidate systems.

Table \ref{data} summarizes our multi-wavelength data set. All 33 fossil group candidates which were observed with the VLA are included; the three candidates which fail the optical criteria with $\Delta$mag $<1.3$ are indicated by an asterisk.  
In the following subsections we present the results of our multi-wavelength comparison between radio, optical, and X-ray characteristics of the central elliptical galaxy.  We investigate scaling relations which may provide insight into the evolution of fossil groups (e.g. \citealt{Khosroshahi07}). 
To understand the radio emission in the context of group evolution and mass assembly we compare our results with a radio investigation of the dominant central elliptical galaxy in X-ray bright, non-fossil groups from the Group Evolution Multiwavelength (GEMS) survey \citep{Osmond04} reported by \citet{Croston05}.


\subsection{$L_X-L_{opt}$: A Consistency Check}

\begin{figure*}[t]
\centering
\includegraphics[width=0.5\textwidth]{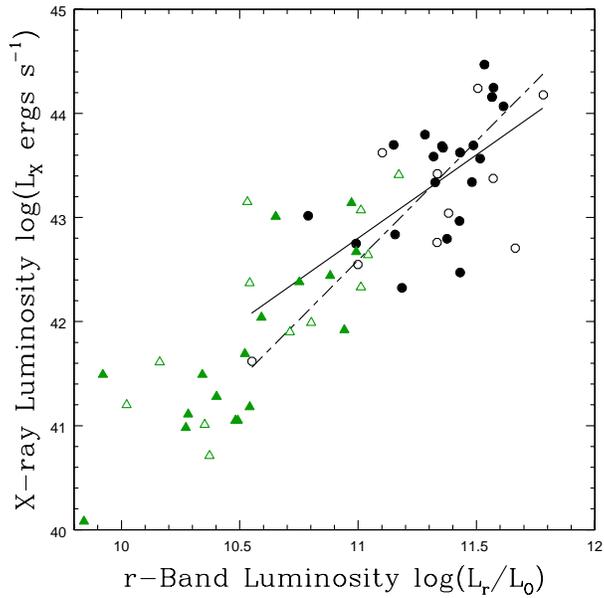}
\caption{The $L_X-L_{opt}$ relation plotted for our sample of fossil groups. Filled circles indicate radio detections, open circles are non-detections for which we calculated 3$\sigma$ upper limits for the presence of a radio source. The solid line is our best fit line to the L$_X$(0.5-2 keV)--L$_{opt}$(\emph{r}-band) fossil group data.  For comparison, the dot-dashed line demonstrates the slope of the best fit from \cite{Jones03} derived from the L$_X$(bolometric)--L$_{opt}$(R-band) data. Green triangles represent the sample of elliptically dominated non-fossil groups from \citet{Croston05} and have not been used in the line fit.}
\label{optxray}
\end{figure*}

\cite{Jones03} and \cite{Khosroshahi07} quantified a relationship between the bolometric X-ray luminosity of the extended hot gaseous halo and the optical R-band luminosity of the central dominant elliptical galaxy of fossil groups.  This correlation is considered indirect evidence that the isolated elliptical galaxy is the result of group merger activity.  In Figure \ref{optxray}, we plot the $0.5-2$ keV luminosity of the extended X-ray halo reported in \citet{Santos07}, versus the SDSS DR7 \emph{r}-band luminosity of the central dominant elliptical galaxy reported in this work.  Filled black circles indicate fossil groups that are detected in our radio survey; open black circles represent fossil groups for which no radio source is detected.  Green symbols are non-fossil, elliptically dominated groups from the GEMS survey \citep{Croston05}.  Interestingly, both detections and non-dections are well distributed over the full range of optical and X-ray luminosities from the brightest to the faintest central elliptical galaxy and fossil group halo.

We conduct a least-squares fit to the fossil group sample and find $\log(L_X)=(1.72\pm0.35)\times\log(L_r)+(23.9\pm4.01)$ fits the data best.  This is difficult to directly compare with previous studies which use the bolometric X-ray luminosity of fossil groups, however we have plotted the slope of the best fit line from \cite{Jones03}, with a slope of $2.29\pm0.24$, in our figure for a relative comparison.  The degree of scatter in the $L_X-L_{opt}$ plot is such that either line could be argued to fit the data.  Our ``best fit'' line is also shallower than that found for a sample of elliptical galaxies ranging from individual elliptical galaxies with $L_X=10^{36}$ erg s$^{-1}$ to cluster galaxies with $L_X=10^{43}$ erg s$^{-1}$ ($2.28\pm0.17$; \citealt{Beuing99}).  Nonetheless, we believe that our sample of fossil groups follows a similar ``scaling relation'' to that of \cite{Jones03} and \cite{Khosroshahi07} and therefore consists of similar systems to fossil groups previously presented in the literature.  Additionally, the central elliptical galaxies are consistent with those in other fossil group samples which are an order of magnitude more luminous in both X-ray and optical luminosities than normal elliptical galaxies, for example compared to the GEMS groups.  Figure \ref{optxray} also shows that the trend between X-ray and optical works for non-fossil groups, extending to dominant elliptical galaxies of lower stellar mass, consistent with previous work (e.g. \citealt{Helsdon03}).

\subsection{$L_X-L_{1.4GHz}$ and $L_{opt}-L_{1.4GHz}$: Multi-wavelength Comparisons}

The 1.4 GHz radio emission is dominated by synchrotron emission from nuclear activity in the central elliptical galaxy, with little to no contribution of free-free emission from star formation.  The presence of radio emission suggests that radio AGN contribute to heating the intragroup medium (IGM) long after the central dominant elliptical galaxy has assembled the majority of its mass.  Two possibilities exist: the AGN may be long-lived or it may be recently rejuvenated.  By connecting the radio properties of the AGN with the X-ray properties of the halo and the optical properties of the central galaxy, we attempt to understand AGN activity within the evolutionary framework of fossil groups.

\begin{figure*}[t]
\centering
\plottwo{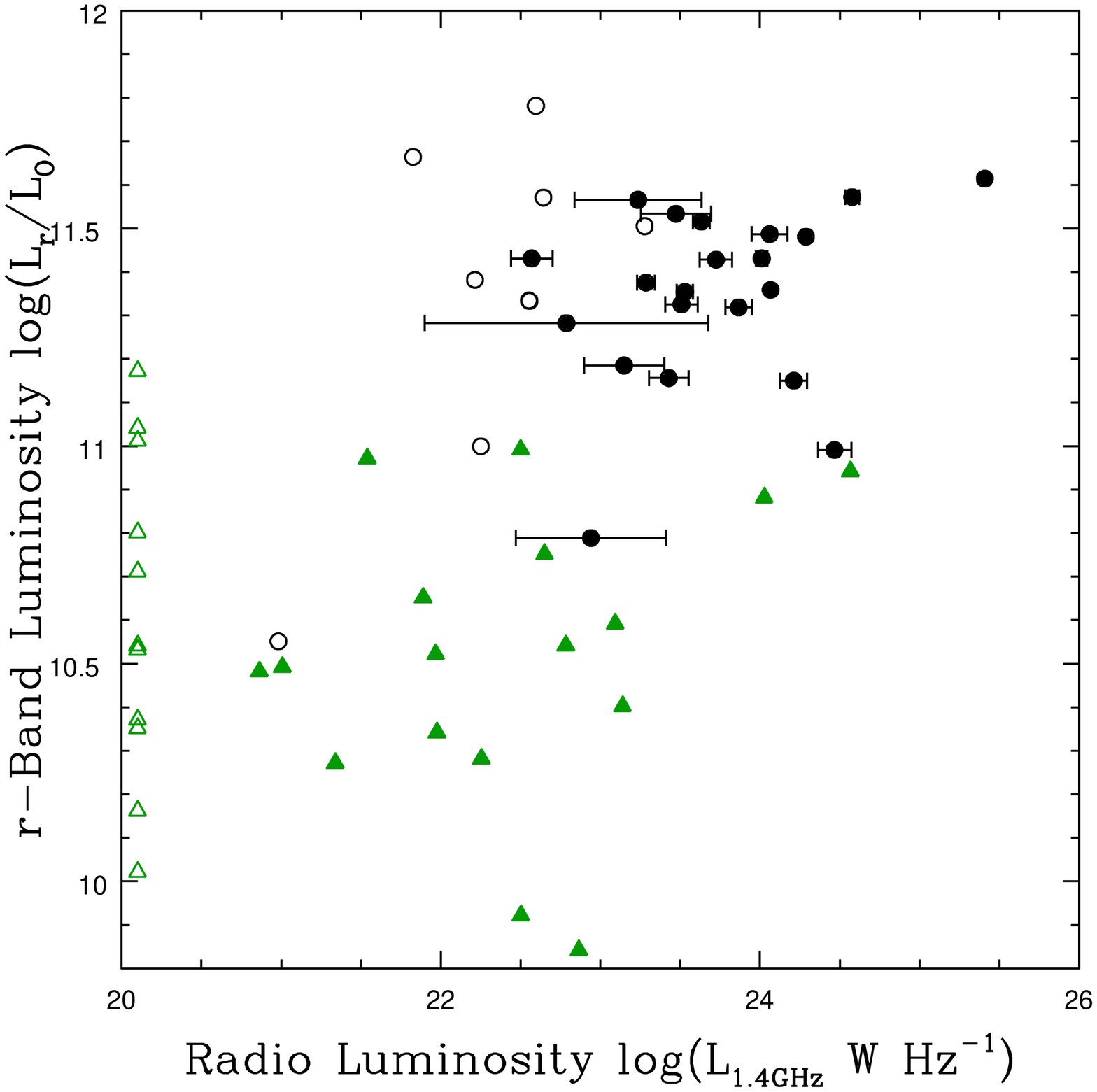}{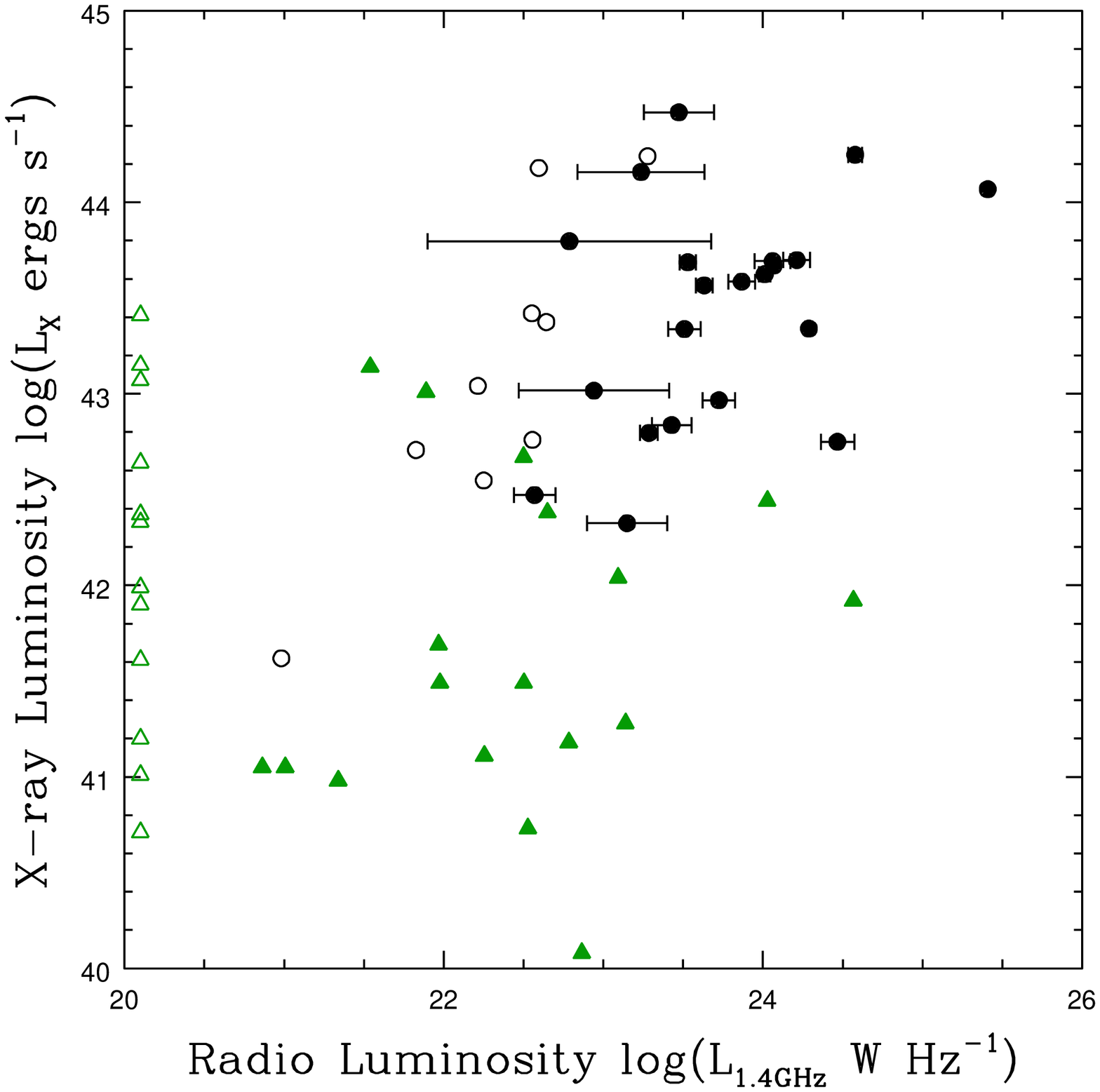}
\caption{Left (a):  Radio luminosity versus \emph{r}-band luminosity of central dominant elliptical galaxy.  Right (b): Radio versus X-ray luminosity.  Filled circles indicate radio detections, open circles are 3$\sigma$ upper limits for those fossil groups for which we do not detect a radio source. Error bars represent 1$\sigma$ \textit{rms} noise from the radio observations.  Green triangles represent the sample of elliptically dominated non-fossil groups from \citet{Croston05}. Non-detections are plotted as open circles to the left to demonstrate their range of optical and X-ray luminosities, respectively.  }
\label{radxray}
\end{figure*}

Figure \ref{radxray}a shows the plot of $r$-band versus radio luminosities.  Figure \ref{radxray}b shows the plot of X-ray versus radio luminosities.   Black points represent our fossil group sample.  Open circles indicate 3$\sigma$ upper limits for non-detections.  Green points represent the sample of elliptically dominated non-fossil groups from \citet{Croston05}.  Non-detections are plotted as open circles to the left to demonstrate their range of optical and X-ray luminosities, respectively.

We find that, within the fossil group population, there is no trend between radio and optical luminosities.  However, when we include the dominant elliptical galaxies in non-fossil groups we find that, in general, the radio emission is an order of magnitude greater in fossil than non-fossil groups.

When we similarly break down the radio--X-ray plot, there is a weak trend among fossil groups that more X-ray luminous objects are also more radio-loud.  This trend is strengthened when we include non-fossil groups.  Both fossil and non-fossil groups that are more luminous in X-rays also tend to be more luminous at radio wavelengths.  This is consistent with the finding of \citet{Croston05} that radio-loud, non-fossil groups are hotter in X-rays than radio-quiet groups, which suggests radio AGN contribute to heating the IGM.

\subsection{$\Delta$mag $-L_{1.4GHz}$: AGN Activity Since Fossil Group Formation}

Fossil groups may only be a phase in the build up of hierarchical structure, however simulations suggest a relationship between the formation time of the fossil group and the difference in optical magnitude between the two brightest group member galaxies ($\Delta$mag; \citealt{D'onghia05,vonBendaBeckmann08}).  The formation time is defined as the epoch at which 50\% of the system's virial mass has been assembled, and the relationship shows that the earlier a fossil group formed, the larger the magnitude gap (Figure 2 of \citealt{D'onghia05}).

\begin{figure*}[t]
\centering
\plottwo{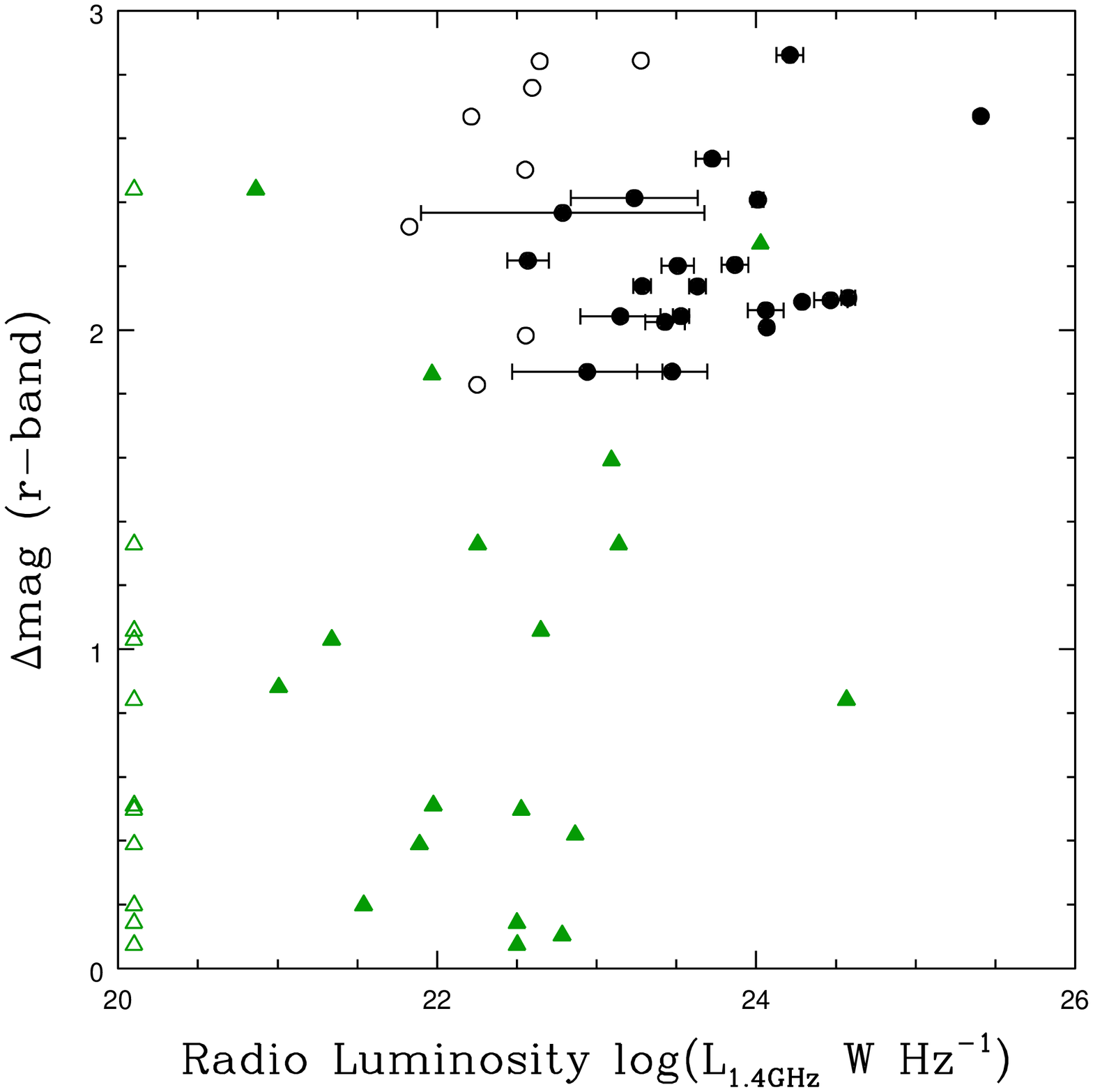}{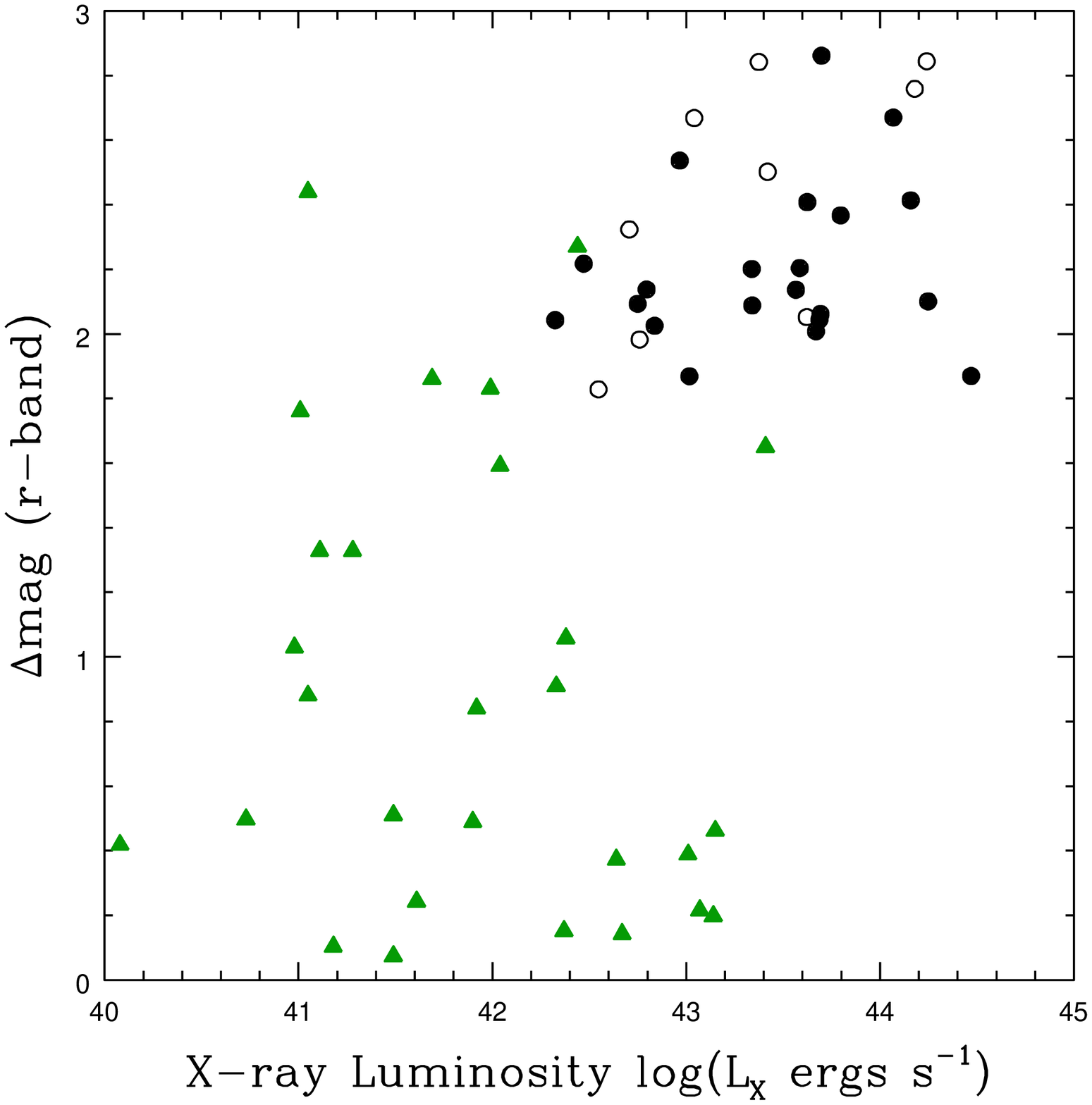}
\caption{Left (a): Radio luminosity versus the difference in magnitude between the first- and second-ranked galaxies.  Right (b): X-ray luminosity versus the difference in magnitude between the first- and second-ranked galaxies.  Symbols are the same as in previous figures.  The optical magnitudes are from DR7, while the sample was originally derived from DR5.  Therefore, the fossil groups below $\Delta$mag$ =2$ are the result of updated photometry.}
\label{deltamag}
\end{figure*}

We plot $\Delta$mag against the radio and X-ray luminosities to investigate a connection between the proxy for the fossil group formation time and the strength of the AGN (Figure \ref{deltamag}a) or the luminosity/temperature of the hot X-ray halo (Figure \ref{deltamag}b).  Again, when we only consider the fossil groups, there is no correlation between the $\Delta$mag and the radio/X-ray luminosities of the central elliptical galaxy.  However, when we include non-fossil groups, we find that groups below the $\Delta$mag $=2$ criterion tend to be fainter in terms of both their radio and X-ray luminosities.  In Figures \ref{deltamag}a and \ref{deltamag}b the green points are the same galaxy groups studied by \citet{Croston05}, but the difference in magnitude between the first- and second-ranked galaxies are reported in \citet{Osmond04}.  We note that the trend between $\Delta$mag and formation time reported by \citet{D'onghia05} is really brought out by the groups with $\Delta$mag $<2$.  Similarly, any trend between $\Delta$mag and radio/X-ray luminosities is only evident when we include non-fossil groups in the comparison.

\section{Discussion} \label{discussion}

\citet{Best04} demonstrated that the preferred environment of radio galaxies is the centers of elliptically dominated groups and poor-to-intermediate richness clusters.  Nonetheless, it is remarkable that we detect the majority of fossil groups as not only radio sources, but as \textit{radio-loud} AGN.  One can no longer consider the otherwise dynamically old systems of fossil groups to be passively or quiescently evolving.  The short lifetime for radio synchrotron emission at 1.4 GHz (a couple hundred Myr after a source has turned off; \citealt{Jones01}) compared to the formation timescale of fossil groups inferred from simulations (several Gyr), or the cooling time of X-ray emitting hot gas suggests that a long-lived or recurring process is driving the AGN activity.  

We argue that the radio luminosity is related to the properties of the group/cluster environment, as measured by the hot X-ray halo, and, less directly, to the mass assembly of the dominant elliptical galaxy, as measured by the optical stellar luminosity.  Indeed, the fact that multi-wavelength trends exist even while some sources remain undetected at 1.4 GHz favors a recurring process which drives AGN activity.  Further, the strength of the radio AGN when it ``turns on'' seems to know something about the physical state of the group/cluster environment and the history of mass assembly: stronger AGN reside in more X-ray luminous halos and more massive central elliptical galaxies.  Finally, the consistency between our results and those of \citet{Croston05} suggest that current AGN activity is linked to the heating of the IGM in fossil groups just as it is in non-fossil groups.

It is not clear to what degree these properties are related to the formation time scale of the fossil group ($\Delta$mag).  This is, first, because the trends seen in Figure \ref{deltamag} are weak, and second, because recent simulations suggest that fossil groups are not as simple as an end point for group evolution.  These structures may continue to undergo mass assembly with the fresh infall of massive galaxies \citep{Dariush10}.  To get an observational handle on this aspect of fossil group evolution, we seek a control sample of non-fossil groups with the same X-ray luminosity and stellar mass distributions as the \citet{Santos07} fossil groups, but spanning the full range of $\Delta$mag.  The challenge in finding an established sample in the literature comes from the requirement of complete multi-wavelength coverage and from the fact that fossil groups occupy the high end of the optical luminosity function \citep{DiazGimenez08,Dariush10,MendezAbreu12}.  Nonetheless, in this section we elaborate on the elliptically dominated subset of the GEMS sample, as well as examine two additional samples from the literature of X-ray bright elliptical and early-type galaxies.  The environment of these galaxies range from loose groups to poor clusters.
In addition, we discuss possibilities for fueling radio AGN activity in fossil groups long after the last major merger has taken place.

\subsection{Comparison Samples}

\citet{Croston05} studied the radio properties of the elliptically dominated galaxy groups selected from the GEMS survey \citep{Osmond04}.  All of these groups contain diffuse X-ray emission measured by ROSAT.  Of these non-fossil groups, 63\% contain radio sources of which 26\% are radio-loud using the $L_{1.4GHz}>10^{23}$ definition.  We converted the B-band luminosities of the dominant elliptical galaxies to SDSS \emph{r}-band, assuming $B-r'=1.32$ \citep{Fukugita95}, and found that they are, on average, an order of magnitude fainter than our fossil groups (e.g. Figure \ref{optxray}).  The distribution of X-ray luminosities is similarly fainter in magnitude compared to the X-ray halos of the fossil groups, making this an interesting comparison sample, but not suitable as a control sample.

\citet{Dunn10} investigated a complete sample of very nearby massive, X-ray bright elliptical and S0 galaxies (all but one galaxy are at distances less than 70 Mpc) across a range of environments.  They find 94\% are detected in radio at some level, and 11\% of their sources are radio-loud.   Again, we converted the B-band luminosities (from \citealt{Beuing99}) and find that they reside an order of magnitude down the luminosity function compared to our fossil groups.

Finally, we compared the optical luminosities of our fossil groups with those of MKW/AWM poor clusters \citep{Morgan75,Albert77}.  We converted V-band magnitudes reported by \citet{Yamagata86} and find that they have the same distribution to our fossil group sample in both optical and X-ray luminosities \citep{Kriss83}.  AWM/MKW poor clusters have a broad range of $\Delta$mag, as great as 1.8, and \citet{Ledlow96} suggests that they may be a precursor to fossil groups.  Thus, MKW/AWM poor clusters may be an excellent comparison sample to represent the non-fossil environment, across a range of $\Delta$mag.  A fraction of these poor clusters was observed at 1.4 GHz by \citet{Burns80,Burns81}, first with single dish observations and followed up with the VLA.  They find that 8\% of the poor clusters are detected above 400 mJy.  This is similar to the fraction of fossil group candidates detected in the NVSS and FIRST surveys (mean \textit{rms} of 450 and 150 mJy, respectively) before our pointed observations.  The AWM/MKW clusters have a closer redshift distribution than our sample of fossil groups.  Nonetheless, the similarity of the brightest cluster/group galaxy between poor clusters and fossil groups suggests that deep radio observations, a complete multi-wavelength investigation, and a comparison with cosmological simulations of galaxy and structure formation may reveal interesting results about their evolutionary state. 

We conclude that the radio/x-ray/stellar mass correlations for galaxies identified as fossil groups are no different than those for non-fossil or poor cluster ellipticals with extended x-ray emission.  More speculatively, our results reinforce the notion that fossil groups are just a phase in the dynamical evolution of groups.
In addition to being an order of magnitude brighter in both $r$-band and X-ray emission, we show that fossil group central elliptical galaxies are also brighter at 1.4 GHz than non-fossil dominant elliptical galaxies.  Our results, combined with those samples above, are consistent with those of \citet{Best05} who find that the fraction of galaxies with a radio-loud AGN increases with increasing stellar mass of the host galaxy.  Even within this small sample of fossil groups, if we bin them by $r$-band stellar luminosity, the fraction of fossil groups detected as radio-loud increases with increasing luminosity.  Although we ignore the complexities of converting $r$-band luminosity to stellar mass, it is interesting to note that the fraction of radio-loud fossil groups as a function of luminosity is comparable at the highest luminosity bin ($\sim$60-75\% at $\log(L/L_{\odot})\sim11.5$) to the fraction of radio-loud brightest cluster galaxies in the highest mass ($\log(M/M_{\odot})\sim11.5$) bin observed by \citet{Best07}.

\subsection{Fueling AGN in fossil groups}

The question pertaining to radio AGN in fossil groups then becomes what is the fueling mechanism which drives AGN activity in these dynamically old systems?  Although optical AGN activity in nearby, predominantly late-type galaxies may be the result of major mergers at late times \citep{Alonso07,Schawinski10}, the nature by which fossil groups formed and their age inferred through simulations rule out major mergers.  The lack of obvious stellar signatures in SDSS images such as shells, tidal tails, arcs, or multiple nuclei in the elliptical isophotes of the central early-type galaxy confirm that they have not undergone a major merger in the last gigayear or more \citep{Schneider82,Thomson90}.  Further, we have examined the SDSS spectra of the dominant elliptical galaxies and found that they are exclusively absorption line spectra, indicating that they do not show signs of optical AGN activity and they have likely not experienced a strong burst of star formation in the last $\sim$20 million years.  In this section we discuss alternative possibilities for fueling radio AGN activity, including cooling flows such as have been invoked in so-called cooling core clusters, minor mergers as may be important in Seyferts, and \hi\ accretion from the surrounding environment as viewed in absorption associated with early-type radio galaxies.

The presence of an AGN may be intricately tied to the thermal state of the IGM in a complex feedback loop with a cooling flow at the center of the group potential.  Modeling of deep X-ray observations of non-fossil groups (e.g. \citealt{Ponman93,David94}) and clusters (e.g. \citealt{Fabian01}) show some evidence for cooling flows, however, the temperature of these systems is not observed to reach the level required for gas to cool to the center of the galaxies harboring cool cores.  Modeling spectroscopic X-ray observations of clusters show that cooling flows only appear to deposit mass at roughly 10\% of the rate expected from simple cooling flow models (e.g. \citealt{David01,Peterson01}).  Nonetheless, cooling flows may be the best candidate for providing fuel to a supermassive blackhole and regulating AGN activity (e.g. \cite{Fabian94} and references therein).

Fossil groups for which the temperature has been measured range from 0.66-4 keV \citep{Khosroshahi07,2004AdSpR..34.2525Y}.  Assuming a density of $n=10^{-3}$ cm$^{-3}$ \citep{Fabian94} and using the cooling function of \cite{2001ApJ...546...63T}, the cooling time of fossil groups is a few gigayears.  The densities are still higher at the center of the group potential, leading to faster cooling times, less than the formation time of simulated fossil groups.  There is evidence in at least one fossil group of a decrease in temperature towards the center of the X-ray profile \citep{Vikhlinin99}, although this could also be evidence of cooler gas originating from the interstellar medium of the central galaxy \citep{1998ApJ...496...73M}.  We note that \citet{Jetha09} are investigating the role of cooling flows in a similar sample of fossil group candidates.

We cannot rule out other possibilities for sustaining an AGN including minor mergers.   It is inconclusive whether minor mergers with dwarf galaxies may be substantial to fuel the central engine in other active galaxies such as Seyferts (e.g. \citealt{DeRobertis98,Corbin00}) and they have yet to be investigated as a fuel source in radio-loud AGN.  This is worth investigating in fossil groups as (1) although Seyferts are identified in the optical, they harbor low luminosity radio AGN and (2) fossil groups are known to have a reservoir of nearby dwarf galaxies as shown by their luminosity functions \citep{MendesdeOliveira06,LaBarbera09}.  The conclusions of \citet{Santos07} and \citet{vonBendaBeckmann08} are that at recent times, the dominant elliptical galaxy in a fossil group grows only through minor mergers.

Alternatively, if \hi\ from primordial origins or past gas rich mergers can exist in the group environment for a long time, accretion of these clouds on to the supermassive blackhole may account for long lived radio AGN activity. \hi\ has been detected in absorption against the radio continuum of compact central objects in a number of early type radio galaxies (e.g. \citealt{Jackson03,Morganti09} and references therein).  Further, \hi\ may be present in emission in up to 50-75\% of all early-type galaxies \citep{vanGorkom97,Morganti06}, and in more than 25\% of nearby radio galaxies \citep{Emonts06}.  These systems may have undergone a major merger several gigayears ago, such that the current episode of radio AGN activity is significantly delayed having started very late in the lifetime of the merger, or else is unrelated to merger completely.  


\section{Conclusions} \label{conclusions}

We present the first systematic survey of fossil galaxy group candidates at radio wavelengths using one of the largest collection of uniformly identified sources in the literature \citep{Santos07}.  With updated photometric and spectroscopic optical data from SDSS DR7, we find 30 of these 33 candidates are fossil groups and that 67\% of them are detected at 1.4 GHz down to an average \emph{rms} noise of $\sim$80 $\mu$Jy.  All but one fossil group detection are radio-loud sources.  This is a large fraction considering fossil groups were believed \emph{a priori} to be old, quiescently evolving galaxy systems.  However, it is consistent with previous studies that correlate a rise in the fraction of radio-loud AGN with a rise in the stellar mass of the host galaxy \citep{Best05}.  We find that MKW/AWM poor clusters \citep{Morgan75,Albert77} may be an excellent sample, spanning a range of $\Delta$mag between the first- and second-ranked galaxies, against which to compare fossil groups to further understand the history of their mass assembly.

The high AGN frequency in fossil groups is most likely not directly related to their unique formation history given the prevalence of radio sources among X-ray selected samples of elliptical galaxies and non-fossil groups \citep{Croston05,Dunn10} and the short radiative lifetime of synchrotron emission at 1.4 GHz compared to the simulated formation time scale of fossil groups \citep{D'onghia05,vonBendaBeckmann08}.  We find a weak trend between the radio and X-ray luminosities of fossil groups, but no trend between the radio and optical properties of the central elliptical galaxy.  However, when we include the brightest group galaxy from a sample of X-ray selected elliptically dominated groups, we find both radio--X-ray and radio--optical correlations such that fossil groups are more luminous at all three wavelengths than non-fossil groups.  Thus, if optical luminosity represents the mass assembly history, and the X-ray halo represents the group-sized dark matter halo in which the galaxy resides, then when an AGN is ``on'', its radio strength is related to the environment in which it resides.

The dominant elliptical galaxy in a fossil group resides at the center of group gravitational potential revealed by the presence of an extended X-ray halo.  In such an environment, the nuclear activity may tied to the thermodynamic state of the hot IGM gas and a cooling flow may both fuel and be regulated by the AGN.   However, there is not yet direct evidence of a cooling flow in these systems.  Deep follow-up imaging of the central galaxies at optical and radio wavelengths is required to exclude the possibility of recent mergers or the existence of nearby cold gas reservoirs that can fuel the central engine.  Finally, fossil groups by definition harbor a population of nearby dwarf galaxies; it is unknown to what degree minor mergers contribute to feeding the black hole and the signatures may be difficult to detect.  It is clear that fossil groups are not only an interesting study of mass assembly in the early Universe, but also as a source of accretion and feedback with the IGM in the present day: their isolation provides a relatively simplified environment in which to study these processes.


\acknowledgments

This work has benefited from useful discussions with J.M.~Stone, L.~Nigra and J.S.~Gallagher, III.  We thank the staff of the Very Large Array and the National Radio Astronomy Observatory who have made these observations possible.  The National Radio Astronomy Observatory is a facility of the National Science Foundation operated under cooperative agreement by Associated Universities, Inc.  Funding for this work was supported by the National Science Foundation through grants AST-0708002 and AST-0506628.

Funding for the SDSS and SDSS-II has been provided by the Alfred P. Sloan Foundation, the Participating Institutions, the National Science Foundation, the U.S. Department of Energy, the National Aeronautics and Space Administration, the Japanese Monbukagakusho, the Max Planck Society, and the Higher Education Funding Council for England. The SDSS Web Site is http://www.sdss.org/.

The SDSS is managed by the Astrophysical Research Consortium for the Participating Institutions. The Participating Institutions are the American Museum of Natural History, Astrophysical Institute Potsdam, University of Basel, University of Cambridge, Case Western Reserve University, University of Chicago, Drexel University, Fermilab, the Institute for Advanced Study, the Japan Participation Group, Johns Hopkins University, the Joint Institute for Nuclear Astrophysics, the Kavli Institute for Particle Astrophysics and Cosmology, the Korean Scientist Group, the Chinese Academy of Sciences (LAMOST), Los Alamos National Laboratory, the Max-Planck-Institute for Astronomy (MPIA), the Max-Planck-Institute for Astrophysics (MPA), New Mexico State University, Ohio State University, University of Pittsburgh, University of Portsmouth, Princeton University, the United States Naval Observatory, and the University of Washington.

\clearpage

\clearpage

\begin{deluxetable}{ccccccccccl}
\rotate
\tablewidth{0pt}
\tabletypesize{\scriptsize}
\tablecaption{Fossil Group Properties}
\tablehead{\colhead{Group}&\colhead{}&\colhead{}&\colhead{\emph{r}-Band}&\colhead{\emph{r}-Band}&\colhead{}&\colhead{L$_X$ (0.5-2 keV)}&\colhead{L$_{1.4GHz}$}&\colhead{1.4 GHz \textit{rms}}&\colhead{Radio Extent}&\colhead{}\\
\colhead{ID\tablenotemark{a}}&\colhead{Name\tablenotemark{a}}&\colhead{z}&\colhead{Mag}&\colhead{Abs Mag}&\colhead{$\Delta$mag}&\colhead{($h_{70}^{-2}$ ergs s$^{-1}$)\tablenotemark{a}}&\colhead{(W Hz$^{-1}$)}&\colhead{($\mu$Jy beam$^{-1}$)}&\colhead{(arcsec)}&\colhead{Notes}}
\startdata
 1\phantom{$\dagger$}&J015021.27-100530.5&0.365&17.26&-24.85&2.41&1.44E+44&2.61E+23& 87 &7.8&1; point source\\
 2\phantom{$\dagger$}&J015241.95+010025.5&0.230&15.72&-24.95&2.76&1.51E+44&(5.96e+22)& 74 &\phm{nothing}&\phm{nothing}\\
 3\phantom{$\dagger$}&J075244.19+455657.3&0.052&14.46&-22.64&2.09&5.62E+42&4.43E+24&177 &57.0&2; large bipolar jets\\
 4$\dagger$&J080730.75+340041.6&0.208&16.38&-24.05&2.41&4.21E+43&1.55E+24& 59 &6.9&3; bipolar jets\\
 5\phantom{$\dagger$}&J084257.55+362159.2&0.282&16.79&-24.52&1.87&2.95E+44&4.50E+23& 72 &8.1&4; extended pt src\\
 6\phantom{$\dagger$}&J084449.07+425642.1&0.054&14.08&-22.97&2.04&2.11E+42&2.13E+23&420 &21.0&5; asymmetric fuzzy\\
 7\phantom{$\dagger$}&J090303.18+273929.3&0.489&18.06&-25.19&2.84&1.74E+44&(2.87e+23)& 79 &\phm{nothing}&\phm{nothing}\\
 8\phantom{$\dagger$}&J094829.04+495506.7&0.409&18.21&-24.22&2.36&6.26E+43&9.28E+22& 48 &\phm{nothing}&\phm{nothing}\\
 9\phantom{$\dagger$}&J104302.57+005418.2&0.125&15.98&-23.16&2.86&4.99E+43&2.47E+24&120 &21&6; background galaxy\\
10\phantom{$\dagger$}&J105452.03+552112.5&0.468&17.69&-25.30&2.67&1.17E+44&3.88E+25& 65 &9.3&7; point source\\
12\phantom{$\dagger$}&J112155.27+104923.2&0.240&16.97&-23.84&2.20&3.85E+43&1.11E+24& 57 &11.0&8; elongated pt src\\
13$\dagger$&J114128.29+055829.5&0.188&16.03&-24.11&2.09&2.19E+43&2.94E+24&119 &9.1&9; extended pt src\\
14\phantom{$\dagger$}&J114647.57+095228.1&0.221&16.36&-24.24&2.06&4.93E+43&1.74E+24& 78 &22.0&10; two fuzzy sources\\
15\phantom{$\dagger$}&J114803.81+565425.6&0.105&16.49&-22.08&1.87&1.04E+43&1.32E+23& 29 &11.7&11; asym jet \& bkgrd QSO\\
16$\ast$&J114915.02+481104.9&0.283&17.60&-23.66&1.03&6.10E+43&(3.89e+22)& 32 &\phm{nothing}&\phm{nothing}\\
17\phantom{$\dagger$}&J124742.07+413137.6&0.155&15.88&-23.70&2.13&6.25E+42&2.92E+23& 58 &6.0&12; point source\\
18\phantom{$\dagger$}&J130009.36+444301.3&0.233&16.88&-23.80&2.50&2.63E+43&(5.39e+22)& 65 &\phm{nothing}&\phm{nothing}\\
19\phantom{$\dagger$}&J133559.98-033129.1&0.177&15.84&-24.14&2.14&3.68E+43&6.48E+23& 64 &8.7&13; point source\\
20\phantom{$\dagger$}&J141004.19+414520.8&0.094&14.66&-23.64&2.22&2.96E+42&5.61E+22& 58 &7.0&14; point source\\
21$\ast$&J144516.86+003934.2&0.306&17.94&-23.65&0.14&7.21E+43&3.32E+24&151 &6.8&15; point source\\
22\phantom{$\dagger$}&J145359.01+482417.1&0.146&15.87&-23.58&1.98&5.75E+42&(5.44e+22)&169 &\phm{nothing}&\phm{nothing}\\
23\phantom{$\dagger$}&J152946.28+440804.2&0.148&15.77&-23.71&2.67&1.10E+43&(2.48e+22)& 74 &\phm{nothing}&\phm{nothing}\\
24\phantom{$\dagger$}&J153344.13+033657.5&0.293&17.23&-24.21&2.00&6.00E+43&(7.29e+22)& 56 &\phm{nothing}&\phm{nothing}\\
25$\dagger$&J153950.78+304303.9&0.097&14.93&-23.50&2.04&4.86E+43&5.12E+23& 86 &10.0&16; extended pt src\\
26$\ast$&J154855.85+085044.3&0.072&13.50&-24.25&1.26&5.09E+42&(1.02e+22)&129 &\phm{nothing}&\phm{nothing}\\
27\phantom{$\dagger$}&J161431.10+264350.3&0.184&15.76&-24.36&2.84&2.37E+43&(6.65e+22)&130 &\phm{nothing}&\phm{nothing}\\
28\phantom{$\dagger$}&J163720.51+411120.2&0.032&14.51&-21.25&4.10&4.16E+41&(1.45e+21)& 93 &\phm{nothing}&\phm{nothing}\\
29$\dagger$&J164702.07+385004.2&0.135&16.13&-23.09&2.02&6.87E+42&4.06E+23& 87 &12.0&17; extended fuzzy\\
30\phantom{$\dagger$}&J171811.93+563956.1&0.114&15.27&-23.55&2.01&4.67E+43&1.77E+24& 81 &11.0&18; extended pt src\\
31\phantom{$\dagger$}&J172010.03+263732.0&0.159&15.44&-24.33&2.10&1.77E+44&5.73E+24& 69 &35.0&19; extended halo\\
32\phantom{$\dagger$}&J172852.16+551640.8&0.148&16.72&-22.81&1.83&3.53E+42&(2.69e+22)& 81 &\phm{nothing}&\phm{nothing}\\
33$\dagger$&J225630.04-003210.8&0.224&16.81&-23.91&2.20&2.18E+43&4.88E+23& 74 &11.0&20; extended pt src\\
34\phantom{$\dagger$}&J235815.10+150543.5&0.178&16.08&-23.98&2.53&9.26E+42&8.02E+23& 85 &16.0&21; small bipolar jets\\
\enddata
\tablenotetext{a}{From \cite{Santos07}}
\tablecomments{$^{\dagger}$Sources previously detected in NVSS or FIRST.
$^{\ast}$Candidates fail optical criteria of fossil groups and are not considered in the analysis.
Column (1) is the group ID number assigned by \citet{Santos07}.   Column (2) is the SDSS name of the dominant elliptical galaxy (and follows the J2000 coordinates).  Column (3) is the spectroscopic redshift.  Column (4) is the SDSS \emph{r}-band apparent magnitude retrieved from SDSS DR7.  Column (5) is the SDSS \emph{r}-band absolute magnitude, assuming $h=0.7$, $\Omega_M=0.3$, and $\Omega_{\lambda}=0.7$. Column (6) is the difference in magnitude between the first and second ranked galaxies.  Column (7) is the X-ray 0.5-2 keV luminosity from ROSAT \citep{Santos07}.  Column (8) is the 1.4 GHz radio luminosity, when detected.  Values in parenthesis are the 3$\sigma$ upper limits for non-detections. Column (9) is the \textit{rms} noise achieved in the individually targeted 1.4 GHz observations presented in this paper.  Column (10) includes notes on the radio detections and the corresponding panel in Figures \ref{images1} and \ref{images2}.
}
\label{data}
\end{deluxetable}
\clearpage

\end{document}